\documentclass[12pt]{article}
\usepackage[T1]{fontenc}
\usepackage{tikz,pgfplots}
\usepackage{multirow}
\usepackage{amsfonts}
\usetikzlibrary{matrix}
\usepgflibrary{shapes.misc}
\usetikzlibrary{arrows,shadows}
\usepackage{graphicx}
\usepackage{bbold}
\usepackage[utf8]{inputenc}
\usepackage{amsmath}
\usepackage{mathtools}
\usepackage{amsfonts}
\usepackage{amssymb}
\usepackage{booktabs}
\usepackage[english]{babel}
\usepackage{setspace}
\usepackage{geometry}
\usepackage{xcolor}
\usepackage{color, colortbl}
\usepackage{tikz}
\newcommand{\bea}{\begin{eqnarray}}
\newcommand{\eea}{\end{eqnarray}}    
\newcommand{\be}{\begin{equation}}
\newcommand{\ee}{\end{equation}}

\newcommand{\diag}{\mathrm{diag}}

\newcommand{\eps}{\varepsilon}
\newcommand{\mtrx}[2]{\left(\begin{array}{#1} #2 \end{array}\right)}

\newcommand{\EE}[1]{\cdot 10^{#1}}
\newcommand{\I}[0]{{\rm i}}

\begin{document}
\thispagestyle{empty}
$\,$

\vspace{32pt}
\begin{center}

\textbf{\Large  Exploring New Physics from $\nu_\tau$ events in OPERA} 

\vspace{30pt}
D. Meloni$^a$
\vspace{16pt}

\textit{$^a$Dipartimento di Matematica e Fisica, 
Universit\`a di Roma Tre\\Via della Vasca Navale 84, 00146 Rome, Italy}\\
\vspace{16pt}

\texttt{davide.meloni@uniroma3.it}
\end{center} 
 \abstract We analyze in details the impact of the $10$ $\nu_\tau$ events seen in the OPERA experiment \cite{Agafonova:2018auq} in constraining the 
 Non Standard Interaction parameter $\eps_{\mu\tau}$ affecting neutrino propagation in matter and the allowed parameter space of models with one sterile 
 neutrino of the $3+1$ type.   
\section{Introduction}

Although the standard neutrino  mixing angles and mass differences  have been determined with
very good accuracy \cite{Esteban:2016qun}-\cite{deSalas:2017kay}, neutrino physics remains an interesting fields where to 
search for non-standard properties beyond those described by the Standard Model (SM).
In order to study the effects of new physics in neutrino oscillation,
it is an useful exercise to look at transition channels not often taken into account in previous analyses, since they can offer 
 an independent check on the bounds already obtained with more traditional approaches. 
Very recently the OPERA Collaboration \cite{Agafonova:2018auq} released the energy spectra of the 10 $\nu_\tau$ events generated via 
the $\nu_\mu \to \nu_\tau$ oscillation from the neutrino beam produced at CERN
and of the related backgrounds, consisting of charm decays to $\tau$ leptons and neutral current events.

In this short paper we attempt to analyse these brand-new data  with the aim of studying their impact in the determination of the Non Standard neutrino Interaction (NSI) parameter
$\eps_{\mu\tau}$ affecting neutrino propagation in matter and the new parameters (angles and mass difference) in the 3+1 sterile neutrino scenario. 
\subsection{The NSI case}
Although it is easy to guess that the limited OPERA statistics cannot improve the already 
stringent bound at 90\% Confidence Level (CL) $|\epsilon_{\mu\tau}| < 0.005$ \cite{Farzan:2017xzy}, it is nonetheless useful to check the 
importance of having a sample of $\tau$ events at our disposal and understand which $\tau$ statistics, efficiencies and energy resolutions 
might be necessary from future experiments 
to contribute in a crucial manner to the search for NP in the neutrino sector.
\noindent

From the analytic point of view, we consider NSI of the form \cite{Roulet:1991sm,Guzzo:1991hi,Bergmann:1999rz}:
\begin{equation}
	\mathcal L_{\rm NSI} = - \frac{G_F}{\sqrt{2}}
	\sum_{\stackrel{f=u,d,e}{a=\pm 1}}
	  \eps^{fa}_{\alpha\beta}[\overline{f} \gamma^\mu (1+a\gamma^5) f]
	  [\overline{\nu_\alpha}\gamma_\mu(1-\gamma^5)\nu_\beta]\,,
\end{equation}
where $f$ is summed over the matter constituents and the parameters
$\eps_{\alpha\beta}^{fa}$  are the entries of a Hermitian matrix $\eps^{fa}$; they give the strength of the NSIs.
As for the derivation of the standard matter effect,
these interactions result in an effective new term 
\begin{equation}
\label{eq:hamnsi}
	H_{\rm NSI} = V \mtrx{ccc}{
		\eps_{ee} & \eps_{e\mu} & \eps_{e\tau} \\
		\eps_{e\mu}^* & \eps_{\mu\mu} & \eps_{\mu\tau} \\
		\eps_{e\tau}^* & \eps_{\mu\tau}^* & \eps_{\tau\tau}}
\end{equation}
that must be added to the neutrino oscillation Hamiltonian in flavor basis, where
$\eps_{\alpha\beta} = \sum_{f,a} \eps_{\alpha\beta}^{fa} N_f/N_e$ and
$V = \sqrt{2}G_F N_e$.
The full three-flavor Hamiltonian describing neutrino propagation in
matter is then given by:
\begin{equation}
\label{eq:fullham}
	H = \frac{1}{2E} U \diag(0,\Delta m_{21}^2,\Delta m_{31}^2) U^\dagger
		+ H_{\rm MSW} + H_{\rm NSI},
\end{equation}
where $U$ is the leptonic mixing matrix, $\Delta m_{ij}^2 =
m_i^2-m_j^2$ and $H_{\rm MSW}$ contains the standard
matter effect. At the typical neutrino energy in OPERA, $E_\nu = {\cal
O}(10)$~GeV, and given that  $L\simeq 732$~km,  where $V$ is
the matter potential $V \simeq 1.1 \EE{-13}$~eV in the Earth's
crust ($\rho \simeq 2.7 \, {\rm g/cm}^3$), both
$\Delta m^2_{31} L / (2 E_\nu) \ll 1$ and $V L \ll 1$.
Thus, neutrino oscillations will not have time to fully develop and the main characteristics of the flavor transition
probabilities will be given by the flavor evolution matrix
$S = \exp(-\I HL)$ at first order in $L$:
\begin{equation}
	S \simeq  1 - \I H L.
\end{equation}
The \emph{off-diagonal} neutrino transition probabilities are then given by
\begin{equation}
	P_{\alpha \beta} = |S_{\beta\alpha}|^2 \simeq |H_{\beta\alpha}L|^2\,,
	\label{eq:interest}
\end{equation}
while, in this expansion,  the dependence on the NSI parameters in the \emph{diagonal} ones 
are obtained by the unitarity condition $P_{\alpha\alpha} = 1 -
\sum_{\beta\neq\alpha} P_{\alpha\beta}$. Thus, the appearance
probability $P_{\mu\tau}$ is mainly affected by
the corresponding NSI element $\eps_{\mu\tau}$, while the survival
probabilities depend on the two off-diagonal NSI elements associated
with the flavor, so that for example $P_{\mu\mu}$ is affected by $\eps_{e\mu}$ and
$\eps_{\mu\tau}$. As expected, the diagonal NSI parameters do not
enter at short baselines. 

With the effects of
$\eps_{\mu\tau}$ included, the leading order transition probability $P_{\mu\tau}$
is given by \cite{Blennow:2008ym}:
\begin{equation}
\label{eq:pmutau}
P_{\mu\tau} = |S_{\tau\mu}|^2 = \left|c_{13}^2\,\sin(2\theta_{23})\frac{\Delta
m_{31}^2}{4E_\nu}+\eps_{\mu\tau}^* V\right|^2 L^2 + {\cal O}(L^3),
\end{equation}
where we have neglected the small mass
squared difference $\Delta m_{21}^2$.

\subsection{The 3+1 sterile case}
We limit ourselves here to the case of Normal Ordering (NO) in the standard three-neutrino sector but we do not impose any restriction on the new mass eigenstate $m_4$, which can then be 
much larger than $m_3$ or of the same order of magnitude of $m_1$. In the first case (and neglecting the effects of the solar mass difference as well as 
any sources of CP violation, including matter), the transition probability $P_{\mu \tau}$ can 
be safely approximated to:
\begin{eqnarray}
 P_{\mu \tau} &\sim& 4 \,U_{\mu 3} U_{\tau 4}  U_{\mu 4} U_{\tau 3} \sin^2 \left[\frac{(\Delta m^2_{41}-\Delta m^2_{31}) L}{4 E_\nu}\right] 
 - \nonumber \\ &&4 \, \left(U_{\mu 1} U_{\tau 3}  U_{\mu 3}  U_{\tau 1} + U_{\mu 2} U_{\tau 3}  U_{\mu 3} U_{\tau 2}\right)
 \sin^2 \left(\frac{\Delta m^2_{31} L}{4 E_\nu}\right) - \label{eq:uno} \\ &&4 \, \nonumber
 U_{\tau 4}  U_{\mu 4}\, ( U_{\mu 1} U_{\tau 1} + U_{\mu 2} U_{\tau 2})\sin^2 \left(\frac{\Delta m^2_{41} L}{4 E_\nu}\right)\,,
\end{eqnarray}
where $U_{ij}$ are the matrix elements of the PMNS matrix enlarged to include the fourth mass eigenstate and $\Delta m^2_{41} = m_4^2 -  m_1 ^2$ 
is the new mass difference of the sterile neutrino model.
In the second case ($m_4 \sim m_1$), expanding for small $\Delta m^2_{41}$ and  $\Delta m^2_{21}$ at first order (but still neglecting CP violating phases and 
matter effects), we get:
\begin{eqnarray}
 P_{\mu \tau} &\sim&
 -4 \, \left[U_{\tau 3} U_{\mu 3} (U_{\mu 1} U_{\tau 1} + U_{\mu 2} U_{\tau 2}) + 
 U_{\mu 3} U_{\tau 4} U_{\mu 4} U_{\tau 3}\right]
 \sin^2 \left(\frac{\Delta m^2_{31} L}{4 E_\nu}\right) +\nonumber  \\ &&4 \,  \label{eq:due} 
  U_{\mu 2} U_{\tau 3} U_{\mu 3} U_{\tau 2}\left(\frac{\Delta m^2_{21} L }{E_\nu}\right) \sin\left(\frac{\Delta m^2_{31}L}{2 E_\nu}\right) 
 +  \\ &&\nonumber
 4\,U_{\mu 3} U_{\tau 4} U_{\mu 4} U_{\tau 3}\left(\frac{\Delta m^2_{41} L }{E_\nu}\right) \sin\left(\frac{\Delta m^2_{31}L}{2 E_\nu}\right)  
\,.
\end{eqnarray}
The relevant feature of these probabilities is that, under the previous hypotheses, the only dependence on the new mixing angles 
is contained in the term $U_{\tau 4} U_{\mu 4}$, to which OPERA can be in principle strongly sensitive, almost independently on the precise value of $\Delta m^2_{41}$. 
In order to analyze the OPERA data, we adopt the following parametrization, particularly useful in the ``atmospheric regime'', 
with oscillations driven by the atmospheric mass difference, $ \Delta m^2_{31} L/ E \sim \pi/2$ \cite{Maltoni:2007zf, Donini:2007yf,Meloni:2010zr}:
\begin{equation}
    \label{eq:3+1param2}
    U =
    R_{34}(\theta_{34}) \; R_{24}(\theta_{24}) \;
    R_{23}(\theta_{23} ,\, \delta_3) \;
    R_{14}(\theta_{14}) \; R_{13}(\theta_{13} ,\, \delta_2) \; 
    R_{12}(\theta_{12} ,\, \delta_1) \,.
\end{equation}
As it is well known, oscillations involving four neutrinos are built with six mixing angles and three CP violating phases; in particular, $\delta_1$ and $\delta_3$ 
are the new phases while $\delta_2$ reduces to the standard $\delta_{CP}$ in the three neutrino case.
According to that, the explicit form of the relevant mixing $U_{\tau 4} U_{\mu 4}$ is given by:
\begin{eqnarray}
\label{tau4}
U_{\tau 4} U_{\mu 4} = \frac{1}{2} \cos ^2\theta_{14} \sin \theta_{24} \sin \theta_{34}\,,
\end{eqnarray}
from which we learn that, for not so large mixings,  relevant changes in $P_{\mu \tau}$ are driven by $\theta_{24}$ and 
$\theta_{34}$; we then expect OPERA to be able to put more stringent bounds on the latter two angles than on $\theta_{14}$.
It has to be noticed that, to our knowledge, a complete fit on the parametrization of eq.(\ref{eq:3+1param2}) is missing 
in the literature; however,  the single matrix elements $U_{\tau 4}$ and $U_{\mu 4}$ are subject to experimental  constraints
\cite{Gariazzo:2017fdh,Dentler:2018sju} which roughly imply $U_{\tau 4} U_{\mu 4} \lesssim 0.04$ at 90\% CL. 
\section{Simulation details and results}
\label{sect:sys}

The OPERA detector \cite{Acquafredda:2009zz} was located in the underground laboratory at Gran Sasso and exposed to the CERN to Gran Sasso 
neutrino beam; fluxes for our numerical simulations have been taken from \cite{flussi} and normalized accordingly to $10^{19}$ 
proton on target (pot) and a detector mass of 1 Kton.
The efficiencies for $\nu_\tau$ identification, as well as the bin-to-bin normalization for both charm and neutral current events, 
have been extracted from the MonteCarlo expectations of Fig.1 of \cite{Agafonova:2018auq}, which refer to the full data sample corresponding 
to $17.97 \cdot 10^{19}$ pot and 
1.25 Kton mass, and the relevant mixing angles fixed to $\theta_{23}=45^\circ$ and $\Delta m^2_{23}=2.50 \times 10^{-3}$ eV$^2$.
For the sake of completeness these numbers are reported in Tab.\ref{tab:events}, 
grouped in 6 energy bins  of variable size in the energy range $E_\nu \in [0, 60]$ GeV; the corresponding total number of 
$\nu_\tau$ events is 6.8, while for the charm and NC backgrounds we have 0.63 and 1.37 events, respectively. 

Instead of a lead target, for the $\tau$ charged current cross section we use an isoscalar target; considering that a 20\% overall systematic uncertainty 
is taken into account 
for the signal error (as well as for the background), we can safely neglect all uncertainties coming from the use of inappropriate cross sections. As for the energy resolution function, we used an energy smearing function 
of a Gaussian form with a standard deviation of the simple type $\sigma (E_\nu) = 0.2\,E_\nu$ \footnote{We checked that a larger value 
$\sigma (E_\nu) = 0.5\,E_\nu$ washes away the OPERA sensitivity to the mixing parameters, almost completely.}.

\begin{table}[h!]
\begin{center}
\begin{tabular}{c  c  c  c}
\toprule
\midrule
 events &   $[0-5]$ GeV & $[5-10]$ GeV &  $[10-15]$ GeV \\ 
\midrule
$\nu_\tau$ app &  0.49 & 2.35 & 2.1 \\
charm back & 0.03 & 0.17 & 0.19 \\
NC back & 0.06 & 0.36 & 0.41 \\
\bottomrule
\midrule
  & $[15-25]$ GeV& $[25-40]$ GeV & $[40-60]$ GeV\\
\midrule
$\nu_\tau$ app &   1.6 & 0.25 & 0.05\\
charm back & 0.18 & 0.04& 0.02\\
NC back & 0.4 & 0.1 & 0.04\\
\bottomrule
\end{tabular}
\caption{\it Number of $\nu_\tau$ appearance (app), charm and neutral current background (back) events expected in OPERA,
corresponding to $17.97 \cdot 10^{19}$ pot and 1.25 Kton mass.
Events are divided in 6 energy bins 
of variable size in the energy range $E_\nu \in [0, 60]$ GeV.}
\label{tab:events}
\end{center}
\end{table}

Our implementation of the $\chi^2$  is based on the pull method \cite{Huber:2002mx,Fogli:2002pt}
and represents the standard implementation of systematic uncertainties in GLoBES \cite{Huber:2004ka,Huber:2007ji}.
For each energy bin $i$ we use a Poissonian $\chi^2$ of the form:
\begin{equation}
  \chi^2 = \sum_i  2 \bigg( F_{i}(\vec{\theta}, \vec{\xi}) - O_{i}
                            + O_{i} \ln \frac{O_{i}}{F_{i}(\vec{\theta}, \vec{\xi}))} \bigg) \,,
\label{equ:chirule}
\end{equation}
where $F_{i}(\vec{\theta}, \vec{\xi})$ is the predicted number of
events in the $i$-th energy bin (for a set of
oscillation parameters $\vec{\theta}$ and nuisance parameters $\vec{\xi}$) and $O_{i}$ is instead the observed
event rate obtained  assuming the true values of the 
oscillation parameters.  Both $F_{i}$ and $O_{i}$ receive
contributions from signal and background rates (indicated with a subscript $s$) specified by $R_{s,i}(\vec{\theta})$, 
so that they can be expressed as: 
\begin{equation}
  F_{i}(\vec{\theta}, \vec{\xi}) = \sum_s \left(1 + a_{s}(\vec{\xi}) \right) R_{s,i}(\vec{\theta}) \,,
\end{equation}
and similarly for $O_{i}$.
The auxiliary parameters $a_{s}$ have the form $a_{s} \equiv \sum_k w_{s,k} \, \xi_k \,,
$ in which the coefficients $w_{s,k}$ specify whether a particular nuisance parameter $\xi_k$ affects the
contribution from the source $s$ or not (so that it assumes the values one or zero, respectively).
Thus, the total $\chi^2_{fin}$, obtained after the minimization over the nuisance parameters $\xi_i$, is given by:
\[
\chi^2_{fin}=\min_\xi\left\{\chi^2
+\sum_k\left(\frac{\xi_{k}}{\sigma_{\xi_k}}\right)^2\right\},
\]
where the last contributions are the pull terms associated with a given systematic parameter $\xi_k$. 
For the sake of simplicity, in our numerical analysis we considered two different sources of systematics, related 
to the overall signal and background normalizations, both fully correlated between different energy bins. As specified above, 
the related uncertainty $\sigma$ is a pessimistic 20\% for both sources. As we will see later, these simple assumptions about
systematics are enough to reproduce the published OPERA results  on the standard 3$\nu$ physics.

The following results are obtained after marginalization over all undisplayed standard and new physics parameters, unless stated otherwise, 
and make use of the full transition probabilities in the standard three neutrino scenario, NSI and 3+1 cases \cite{Kopp:2006wp}.
Notice that for the central values and relative uncertainties of the standard mixing angles and mass differences 
we adopt the latest results in \cite{Esteban:2016qun}, see Tab.(\ref{bestfit}), but for  the leptonic CP 
phase $\delta_{CP}$ which is left free in $[0,2\pi)$. 

\begin{table}
\begin{center}
\begin{tabular}{c  c  c}
\toprule
\midrule
parameter & central value ($^\circ$)& relative uncertainty \\
\midrule
$\theta_{12}$ & 33.62 & 2.3\% \\ 
$\theta_{23}$ (NH) & 47.2  & 4.0\% \\ 
$\theta_{23}$ (IH) & 48.1  & 3.6\% \\ 
$\theta_{13}$ & 8.54  & 1.8\% \\ 
$\Delta m^2_{21}$ & 7.4$\times10^{-5}$~eV$^2$ & 2.8\% \\ 
$\Delta m^2_{31}$ (NH) & 2.49$\times10^{-3}$~eV$^2$ &  1.3\% \\ 
$\Delta m^2_{31}$ (IH) & -2.46$\times10^{-3}$~eV$^2$ &  1.3\% \\
\midrule
\bottomrule
\end{tabular}
\caption{\it Central values and relative uncertainties of the standard mixing parameters extracted from \cite{Esteban:2016qun}. 
For non-Gaussian parameters, the relative 
 uncertainty is computed using 1/6 of the 3$\sigma$ allowed range.}
\label{bestfit}
\end{center}
\end{table}

\section{Numerical results}   
\label{sect:numres}
\subsection{The case of NSI}
\begin{figure}[h!]
\begin{center}
  \includegraphics[scale=0.49]{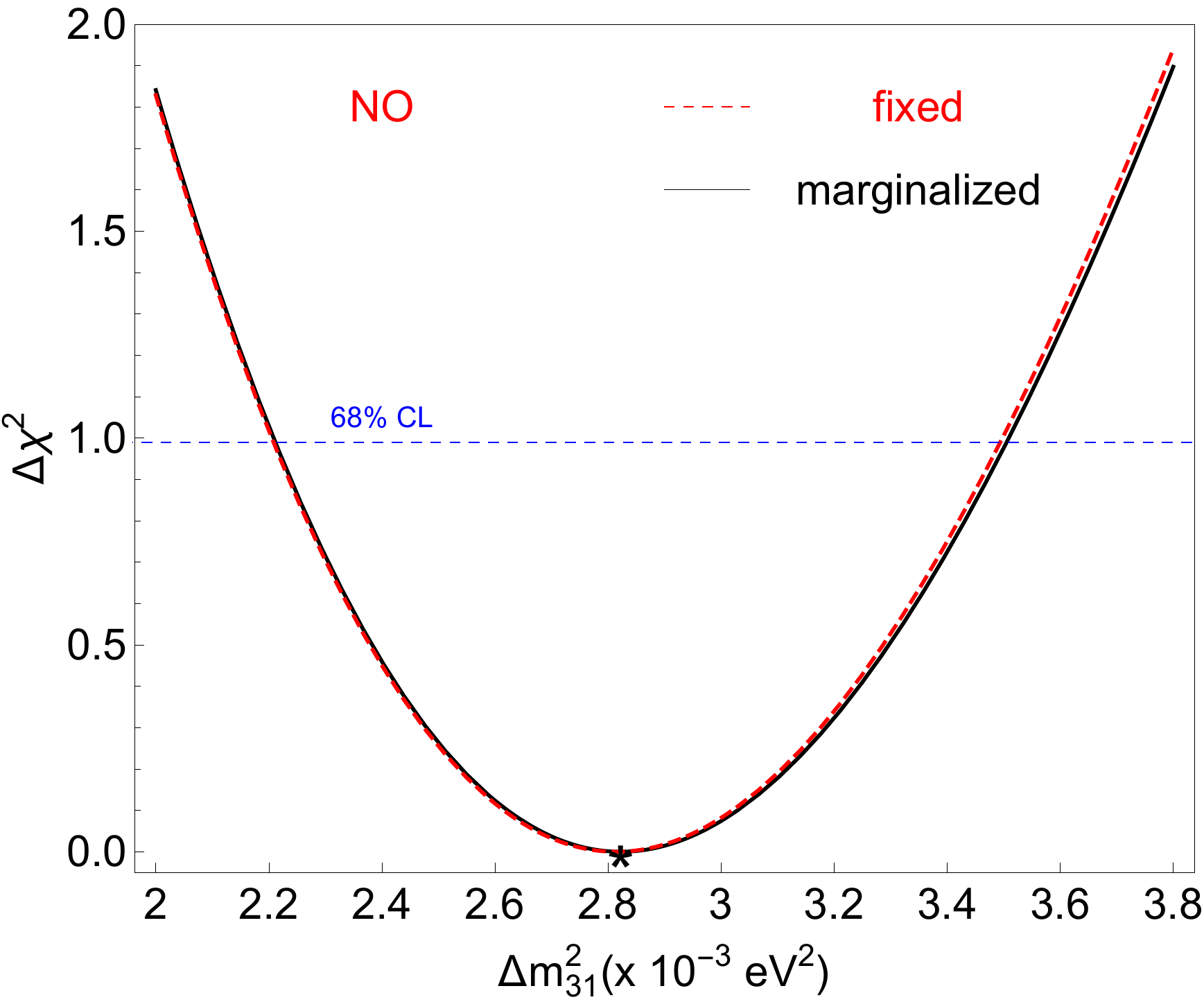}  \includegraphics[scale=0.49]{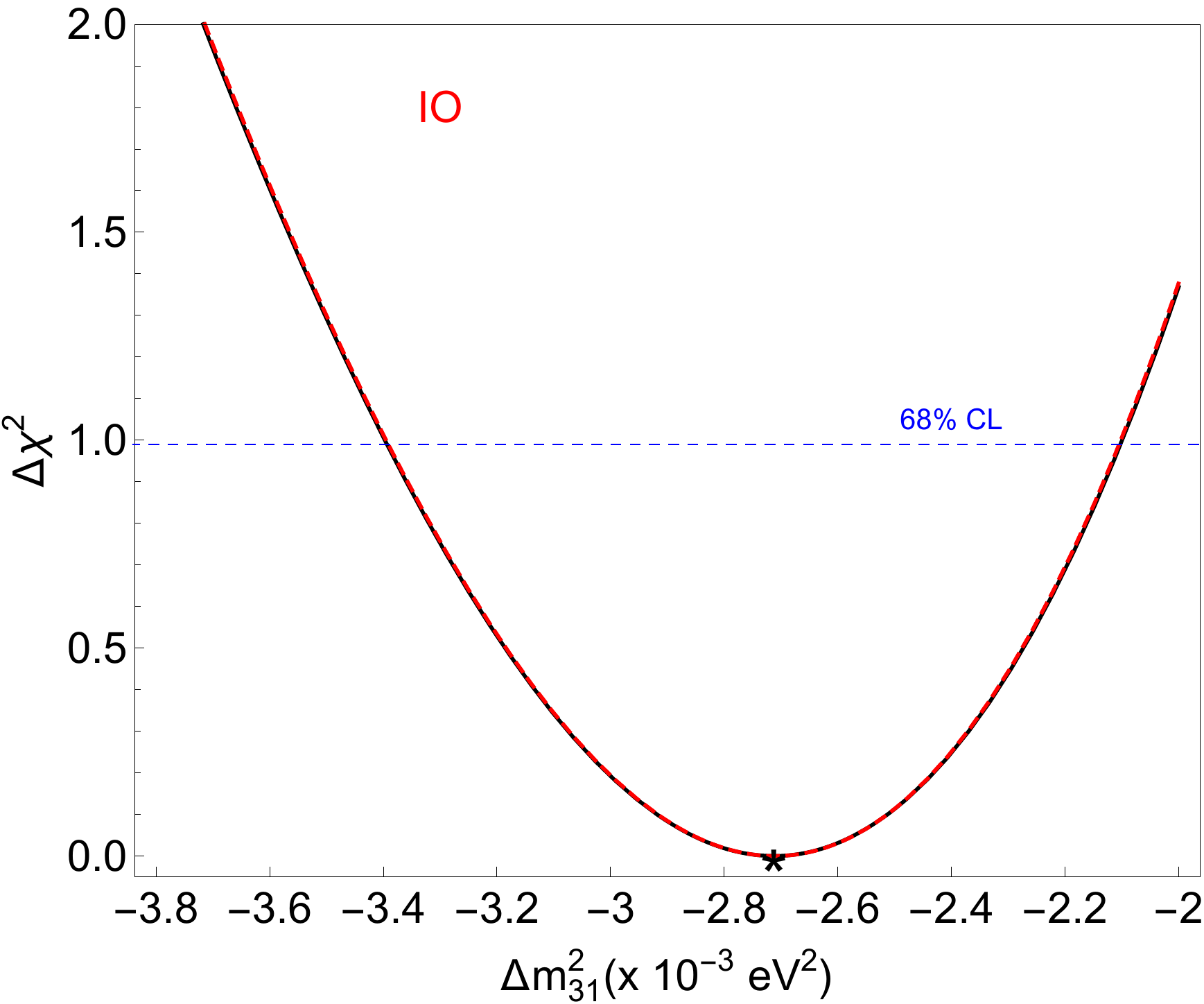} 
\caption{\it \label{fig:dm31} $\Delta \chi^2=\chi^2 - \chi^2_{min}$ as a function of the true $\Delta m^2_{31}$ for both Normal (NO - left panel) 
and Inverted (IO - right panel) neutrino mass orderings. In each panel two cases have been reported: where all standard oscillation 
parameters are kept fixed during the minimization procedure (red dashed line) and when they are all marginalized 
(black solid line). Stars represent the best fit points obtained from our fit. }
\end{center}
\end{figure}

Before discussing in details the bounds on $\eps_{\mu\tau}$, it is useful to reproduce the results on {\it standard physics} quoted by the 
OPERA Collaboration, in particular, the bounds obtained for the atmospheric mass difference $\Delta m^2_{31}$. This is shown in Fig.(\ref{fig:dm31}) 
where we reported the variable $\Delta \chi^2 = \chi^2 - \chi^2_{min}$ as a function of the true $\Delta m^2_{31}$. We have analyzed the two cases 
where all standard oscillation 
parameters are kept fixed during the minimization procedure (red dashed line, {\it fixed} in the legend) and when they are all marginalized 
(black solid line, {\it marginalized} in the legend). The exercise is repeated for both Normal (NO, left panel of Fig.(\ref{fig:dm31})) 
and Inverted (IO, right panel of Fig.(\ref{fig:dm31})) orderings of the 
neutrino mass eigenstates. 

As expected, given the small number of events, not a huge difference can be appreciated when marginalizing over the standard parameters, 
only a modest improvement is seen at large $\Delta m^2_{31}$ for the NO case.
The value of the mass differences and their 68\% CLs (at 1 degree of freedom) 
obtained from our fit are:
\begin{eqnarray}
(\Delta m^2_{31})_{NO} =2.8^{+0.7}_{-0.6} \times 10^{-3} \, {\rm eV^2} \qquad (\Delta m^2_{31})_{IO} =-2.7^{+0.7}_{-0.6} \times 10^{-3}\, {\rm eV^2} \,,
\end{eqnarray}
largely compatible with the OPERA result $|\Delta m^2_{32}| =2.7^{+0.7}_{-0.6} \times 10^{-3}$ eV$^2$ (obtained under the 
assumption of $\sin^2 (2\theta_{23})=1$), thus signaling a good implementation of the experimental systematics in our 
numerical analysis.

We next analyze the bounds on $\eps_{\mu\tau}$ from the $\tau$ appearance data and some interesting correlations with standard parameters (only 
NO is considered in the following). 

\begin{figure}[h!]
\begin{center}
  \includegraphics[scale=0.50]{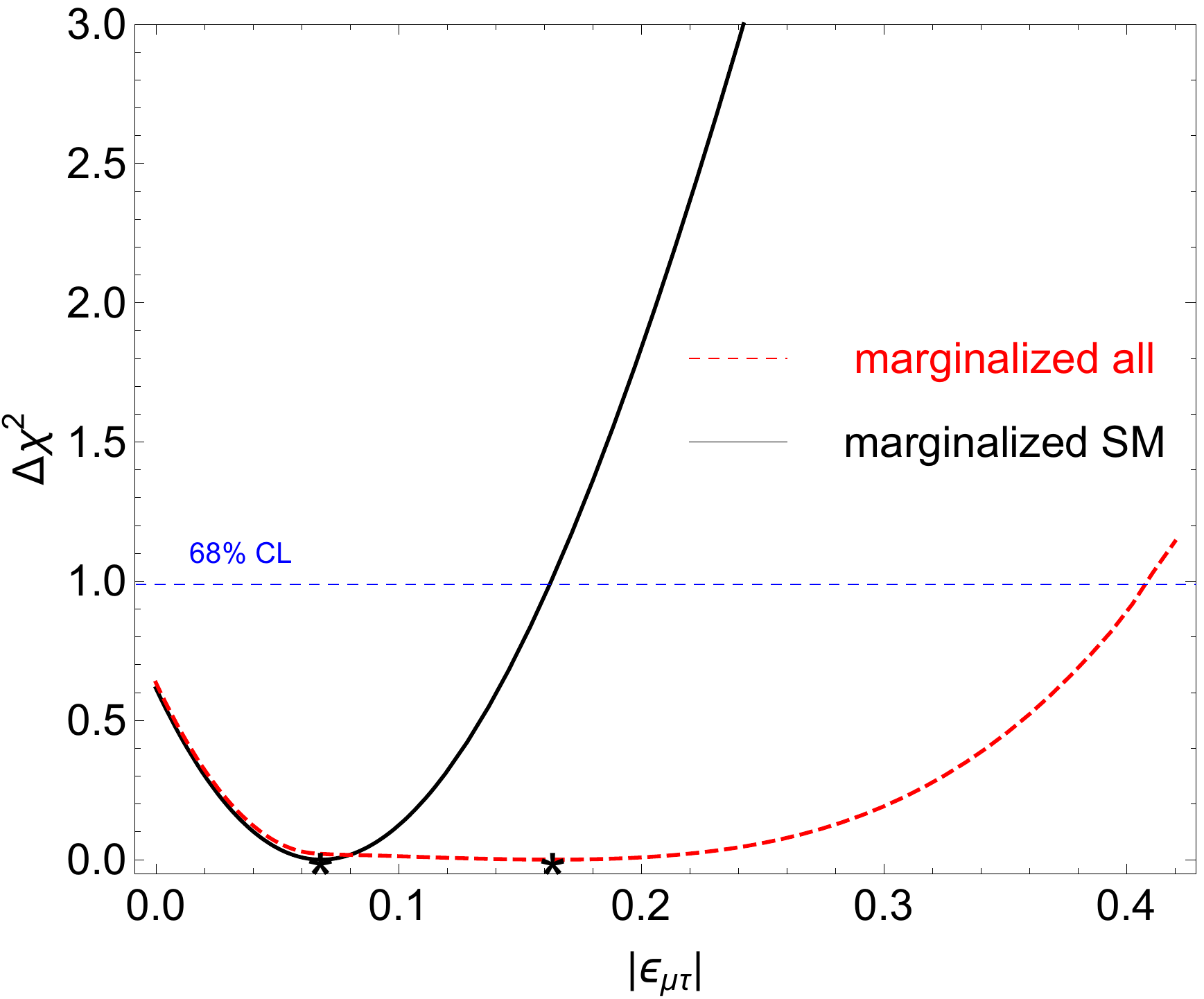}  
\caption{\it \label{fig:abseps} $\Delta \chi^2 = \chi^2 - \chi^2_{min}$ as a function of the true $|\eps_{\mu\tau}|$. 
The solid black line is the results obtained when marginalization is performed on the Standard Model parameters and all NSI parameters are set to zero;
the red dashed line, instead, is computed marginalizing over all oscillation parameters. Stars represent the best fit points obtained from our fit. }
\end{center}
\end{figure}

In Fig.(\ref{fig:abseps}) we report the behavior of the $\Delta \chi^2 = \chi^2 - \chi^2_{min}$ as a function of the true $|\eps_{\mu\tau}|$; 
we perform the fit using two different approaches for the minimization procedure: with the solid black line we represent the results obtained 
when marginalization is performed on the Standard Model parameters and all NSI parameters are set to zero, while 
the red dashed line is obtained marginalizing over all oscillation parameters, including the NSI ones.
As expected, we see that the 68\% CLs strongly depend on the chosen minimization procedure; the upper bounds in the two cases are the following:
\begin{eqnarray}
 |\eps_{\mu\tau}|^{SM} < 0.16 \qquad |\eps_{\mu\tau}|^{all} < 0.41 \,,
\end{eqnarray}
where the superscripts $SM$ and $all$ refer to the results obtained 
when marginalization is performed on the Standard Model parameters and over all oscillation parameters, respectively.
It is interesting to observe that the bound $all$ is roughly of the same order of magnitude as the one set by DUNE \cite{Coloma:2015kiu,Meloni:2018xnk}.

\begin{figure}[h!]
\begin{center}
  \includegraphics[scale=0.44]{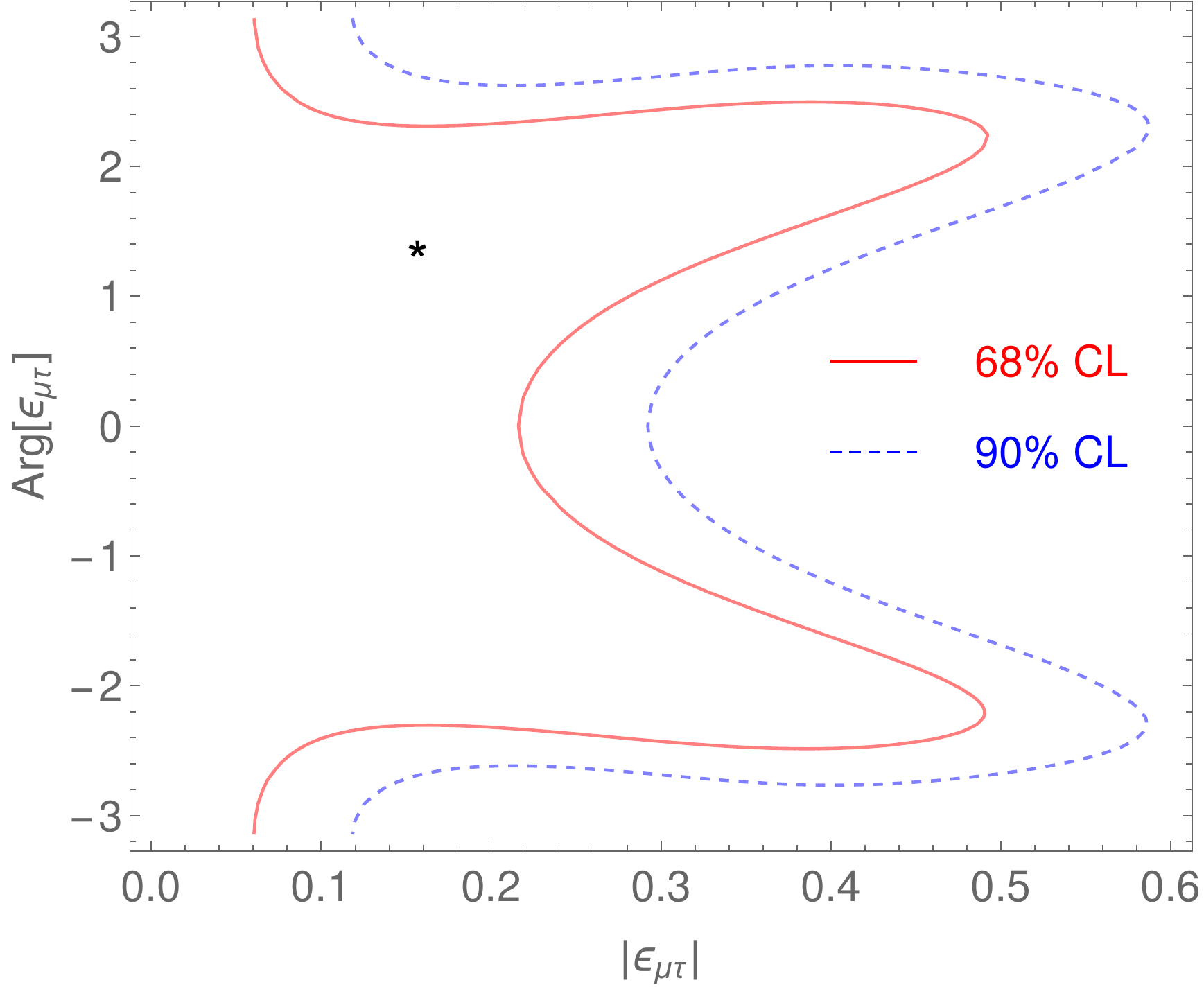}  \qquad \includegraphics[scale=0.48]{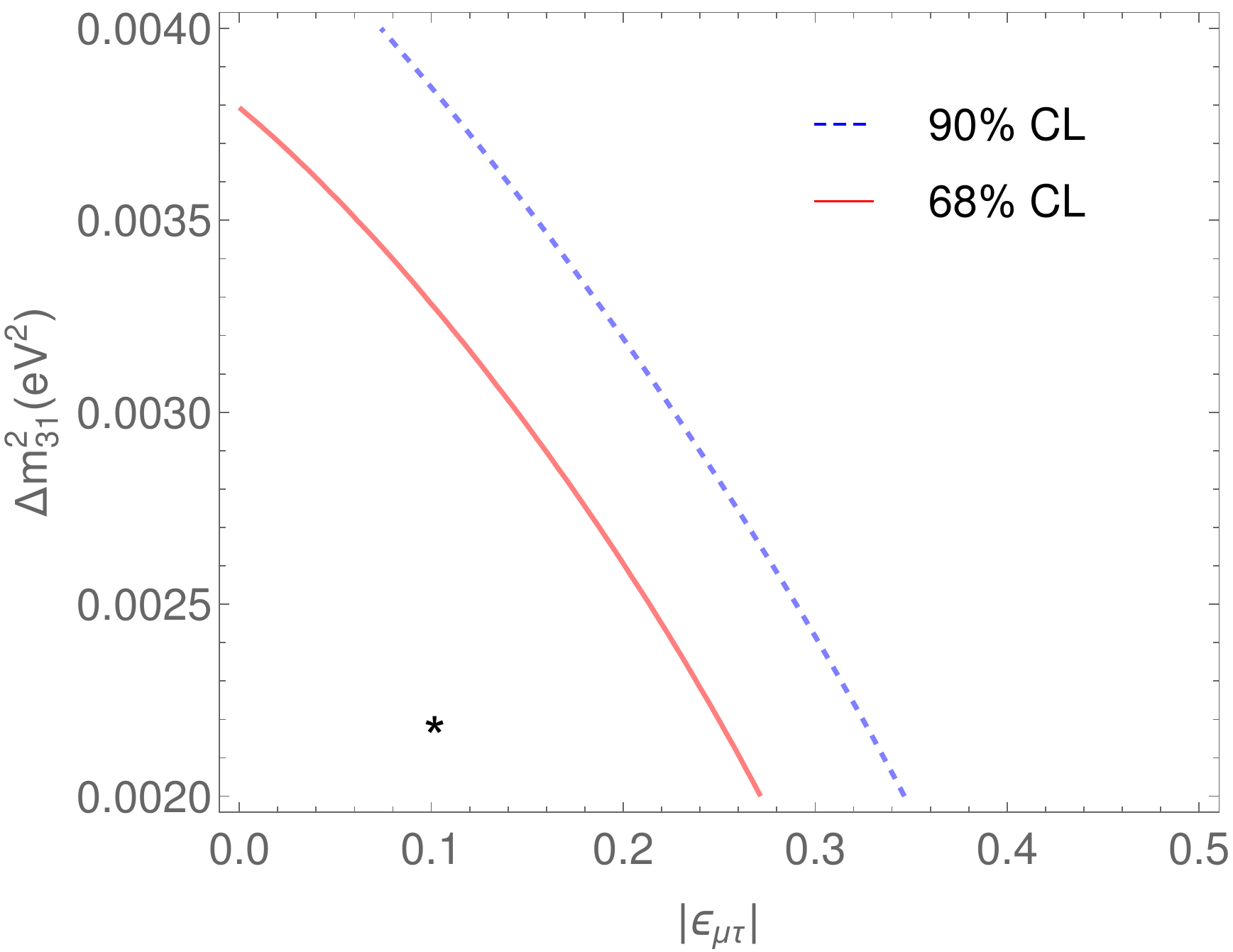} 
\caption{\it \label{fig:absarg} Left panel: correlations between $|\eps_{\mu\tau}|$ and its CP phase 
$\arg(\eps_{\mu\tau})$. Undisplayed parameters have been marginalized over. 
Right panel: correlation between $|\eps_{\mu\tau}|$ and $\Delta m^2_{31}$. Marginalization has been performed on the SM parameters only.
 In both panels we present the 68\% (red solid line) and 90\% CLs (blue dashed line).}
\end{center}
\end{figure}

Finally, in Fig.(\ref{fig:absarg}), we present two potentially interesting correlations between $|\eps_{\mu\tau}|$ and its CP phase 
$\arg(\eps_{\mu\tau})$ (left panel) and $|\eps_{\mu\tau}|$ and $\Delta m^2_{31}$ (right panel). Shown are the 68\% (red solid line) and 
90\% CLs (blue dashed line) for both cases.

As we can see in the left plot, the determination of $|\eps_{\mu\tau}|$ strongly depends on the assumed value of its CP phase,  
the more stringent determination being reached at $\arg(\eps_{\mu\tau})\sim \pm \pi$, with a best fit in the point (indicated with a black star)
$(|\eps_{\mu\tau}|,\arg(\eps_{\mu\tau}))=(0.16,1.40)$. This means that, to maintain the value of $P_{\mu\tau}$ almost constant, see eq.(\ref{eq:pmutau}), 
the regions closed to the CP conserving cases $\exp[i\arg(\eps_{\mu\tau})]\sim \pm 1$ must prefer smaller values for $|\eps_{\mu\tau}|$, as shown in the 
figure.

In the right plot the correlation between $|\eps_{\mu\tau}|$ and $\Delta m^2_{31}$ does not appear to be really significant; we obtained a best 
fit point in $(|\eps_{\mu\tau}|,\Delta m^2_{31})=(0.10,2.2\times 10^{-3}$ eV$^2)$.

\subsection{The case of sterile neutrinos}
\begin{figure}[h!]
\begin{center}
  \includegraphics[scale=0.32]{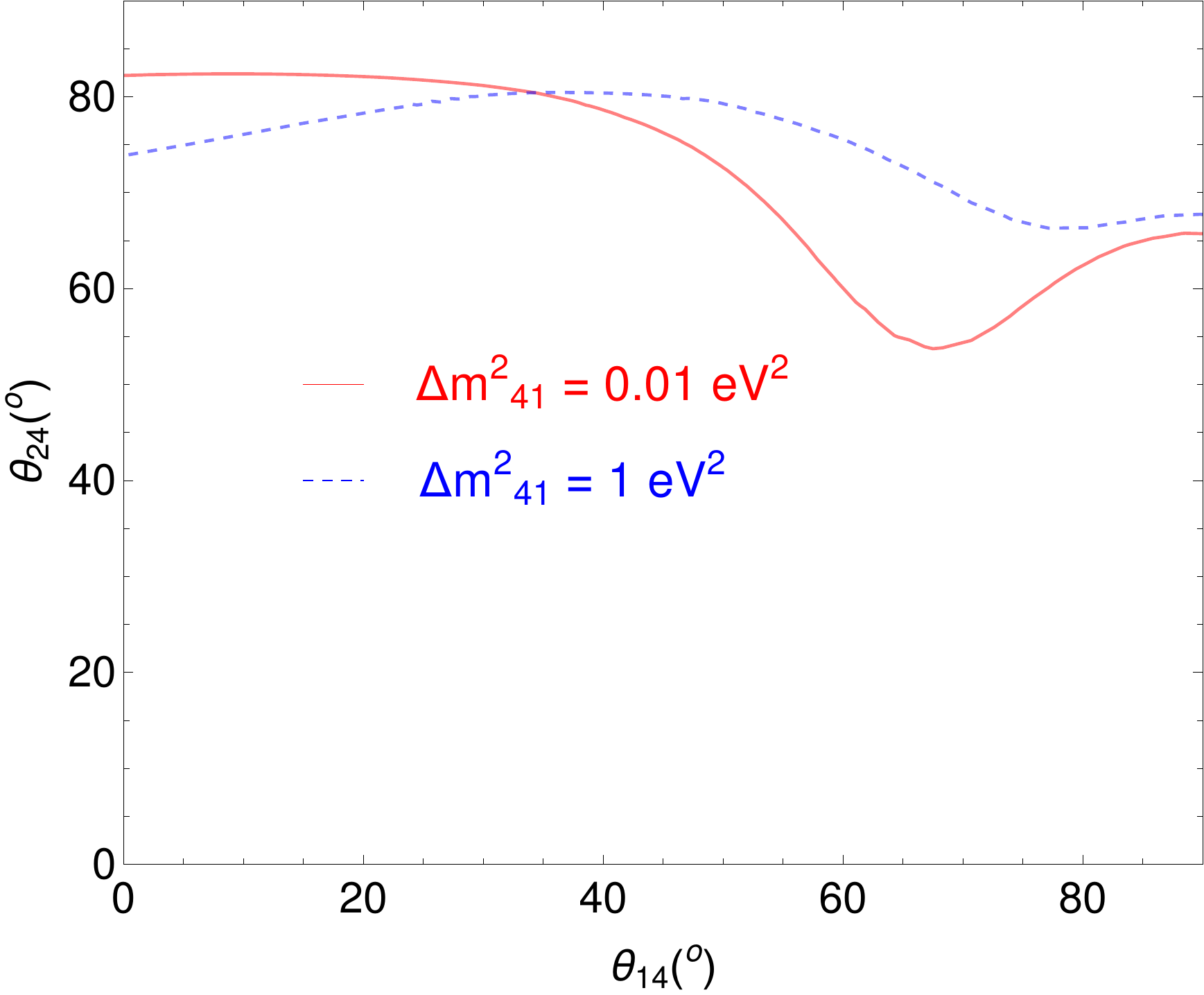} \,  \includegraphics[scale=0.32]{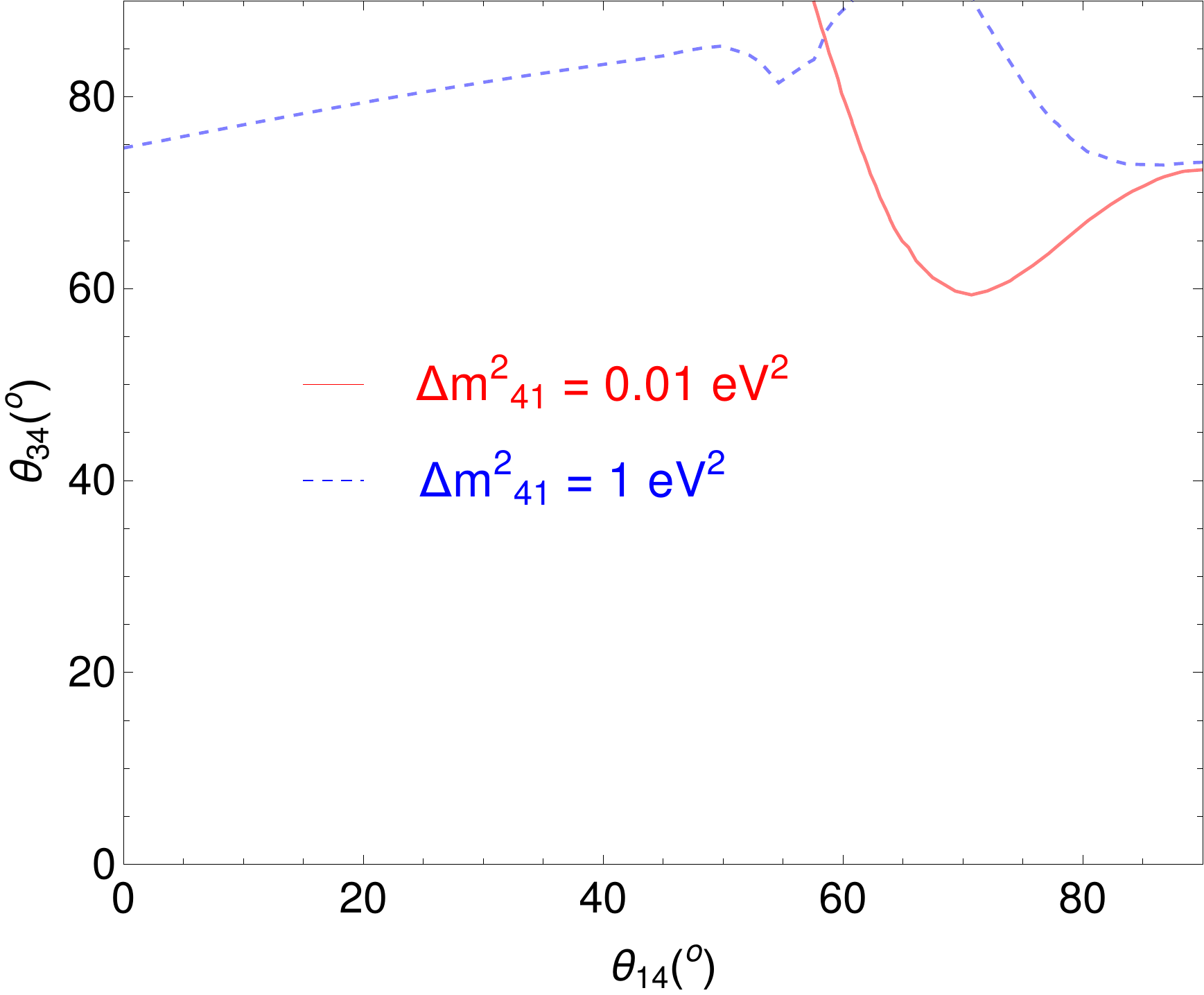}  \,\includegraphics[scale=0.32]{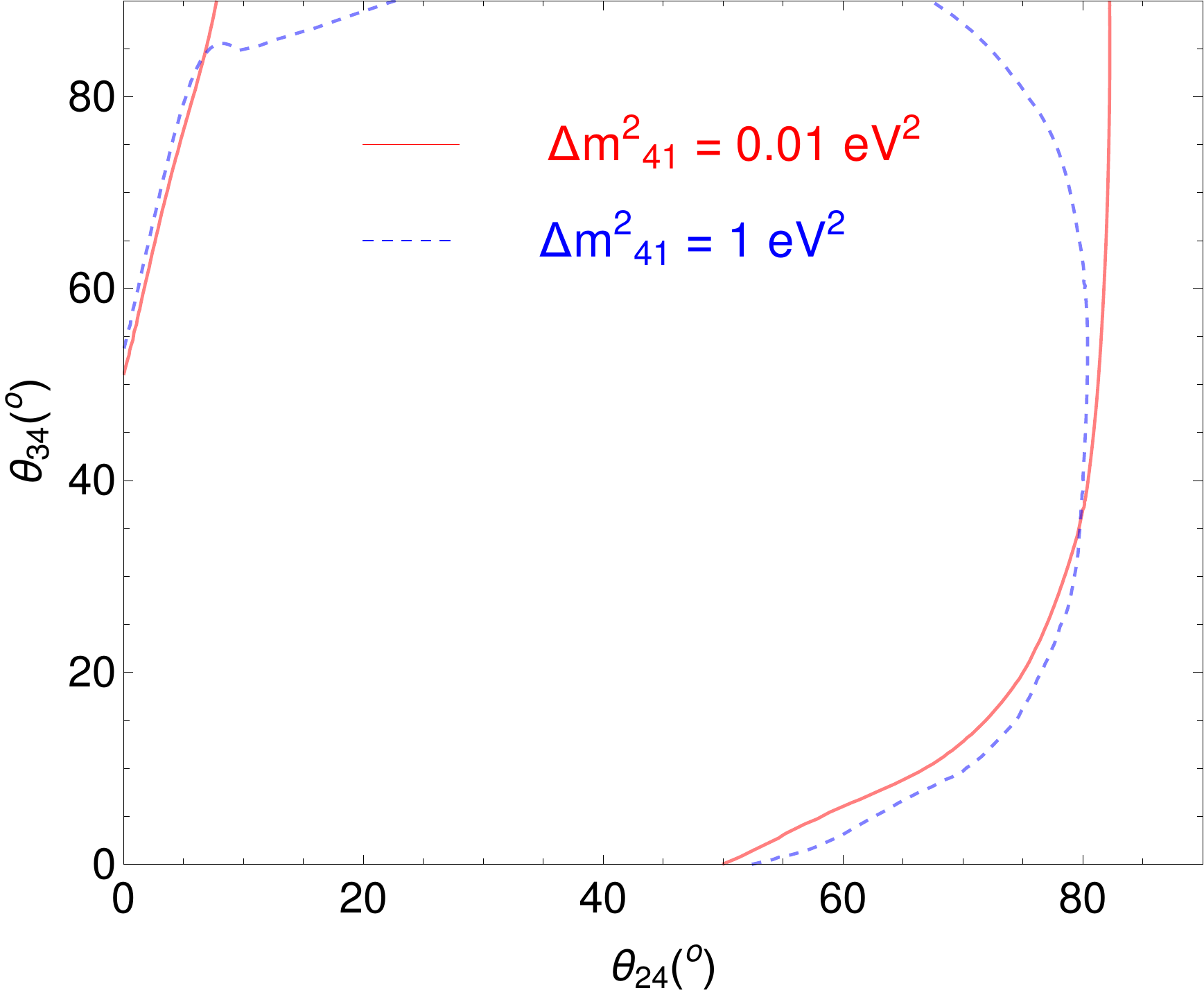} 
\caption{\it \label{fig:angles} 68\% CL bounds on the new mixing angles $\theta_{i4}$ as obtained from the OPERA data. 
Solid red lines refer to the case where $\Delta m^2_{41}=0.01$ eV$^2$, blue dashed lines to $\Delta m^2_{41}=1$ eV$^2$.}
\end{center}
\end{figure}

In the case of the $3+1$ scheme, we start  presenting in Fig.(\ref{fig:angles}) the 68\% CL bounds on the new mixing angles $\theta_{i4}$ in the planes 
($\theta_{14},\theta_{24}$)-left panel, 
($\theta_{14},\theta_{34}$)-central panel and  ($\theta_{24},\theta_{34}$)-right panel.
We fixed the new mass difference to two distinct values 
$\Delta m^2_{41}=0.01$ eV$^2$ (solid red lines) and $\Delta m^2_{41}=1$ eV$^2$ (dashed blue lines).

\begin{figure}[h!]
\begin{center}
  \includegraphics[scale=0.5]{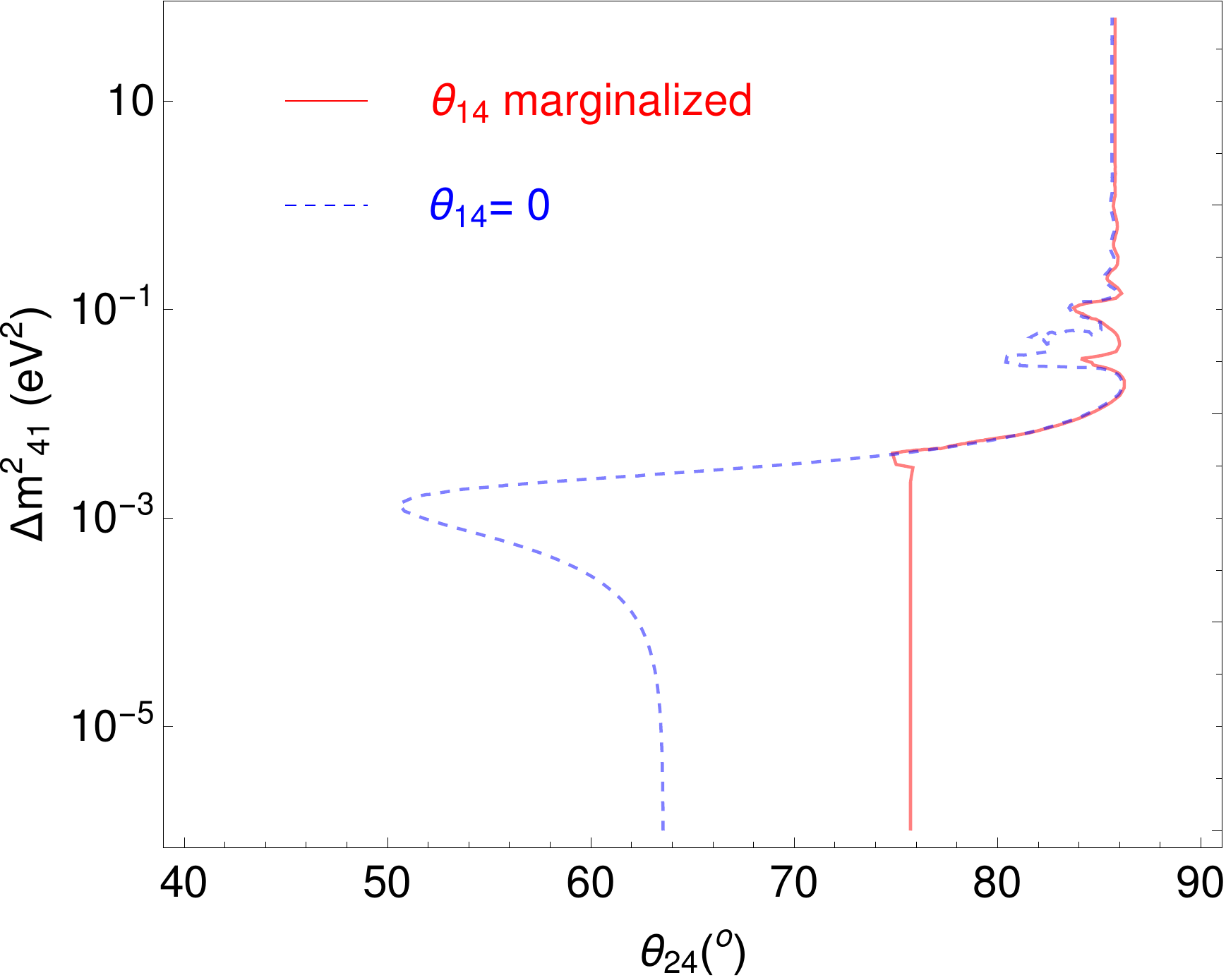} 
\caption{\it \label{fig:dmsq}  90 \% CL excluded region in the
$\left(\theta_{24},\Delta m^2_{41}\right)$-plane for the two cases where $\theta_{14}$ is fixed to be vanishing (blue dashed line) and where 
$\theta_{14}$ is marginalized in the whole $[0,\pi/2]$ interval (red solid line).}
\end{center}
\end{figure}

The results corresponding to the two sets of mass differences do not show  very different  bounds on the mixing
angles and, in particular, no appreciable limits can be put on $\theta_{14}$, as remarked 
after eq.(\ref{tau4}); limits on $\theta_{24}$ and $\theta_{34}$, instead, are of the same order of magnitude, $\theta \lesssim 60^\circ$ at the best. 
However, the importance of $\theta_{14}$ is more visible when addressing the sensitivity of OPERA to the new mass scale; this is because, in the 
marginalization procedure, the $\chi^2$ can be minimized for very large $\theta_{14}$, thus destroying the good sensitivity obtainable 
when the angle is fixed to a vanishing value. To illustrate this point, we present in Fig.(\ref{fig:dmsq}) the 90 \% CL excluded region in the
$\left[(\theta_{24},\Delta m^2_{41})\right]$-plane for the two cases where $\theta_{14}$ is fixed to be vanishing (blue dashed line) and where 
$\theta_{14}$ is marginalized in the whole $[0,\pi/2]$ interval (red solid line).

As it can be seen, the largest difference among the two cases appears for values of the new mass difference smaller than the atmospheric 
$\Delta m^2_{31}\sim 10^{-3}$ eV$^2$, where the first term in eq.(\ref{eq:due}) dominates and large $\theta_{14}$ can cancel the contributions 
from the standard mixing, thus reducing the sensitivity of the experiment to $\theta_{24}$. 

The pattern seen in the figure, independently on the adopted strategy for the marginalization, is clear: 
in the region $10^{-4} \,eV^2 \lesssim \Delta m^2_{41} \lesssim \sim 10^{-2} \,eV^2$, the new mass difference is of the same order as the 
atmospheric frequency in NO, so that  interference effects lead to a sensitive 
improvement in the exclusion regions down to $\lesssim 50^\circ$; on the other hand, for smaller values of $\Delta m^2_{41}$, 
the active-sterile oscillations due to the new frequency are suppressed but, being $\Delta m^2_{43}\sim -\Delta m^2_{31} $, 
the oscillations due to this new mass difference  continue to be present, thus justifying a non-vanishing sensitivity to the mixing angle,
almost irrespective on the precise $\Delta m^2_{41}$ value.

\section{Conclusions}
In this paper we have addressed the possibility of studying two common new physics scenarios in neutrino oscillations, namely NSI and $3+1$ 
sterile neutrino mixing, using the recent published data of the OPERA experiment on the $\nu_\mu \to \nu_\tau$ transition. Although the statistics
at our disposal is not sufficiently large as to expect huge improvements in the bounds already set on the new physics parameters, we nonetheless 
considered this exercise as an interesting one because  the contributions of the new $\nu_\tau$ appearance data have been explicitly taken 
into account.

In the NSI sector, the relevant new parameter to which OPERA is maximally sensitive is $\eps_{\mu\tau}$ on which, using the spectral informations 
released in \cite{Agafonova:2018auq}, can set the 90\% CL upper bound $|\eps_{\mu\tau}|< 0.41$, roughly 
two orders of magnitude worse than the current constraint.
Anything can be said on the related CP phase, which remains undetermined in the whole $[0,2 \pi)$ interval.

In the $3+1$ sterile neutrino case, the OPERA data showed a limited sensitivity only on the correlated $\theta_{24,34}$ angles (in the adopted parametrization 
of the PMNS given in eq.(\ref{eq:3+1param2})):  values larger than $\sim 60^\circ$ can be excluded if one of the other angle is close to 
be vanishing, for almost any value of the new mass difference $\Delta m^2_{41}$ in the range $[10^{-6},10^2]$ eV$^2$.

\section*{Acknowledgments}
I am grateful to Anish Ghoshal for the careful reading of the manuscript.

\bibliographystyle{MLA}

\end{document}